\newcounter{myctr}
\def\myitem{\refstepcounter{myctr}\bibfont\noindent\ifnum\themyctr>9\else\phantom{0}\fi\hangindent17pt\themyctr.\enskip}

\documentclass{ws-ijqi}
\usepackage[super,sort,compress]{cite}
\usepackage{url}

\usepackage{graphicx}
\usepackage{dcolumn}
\usepackage{bm}
\usepackage[mathlines]{lineno}

\usepackage{newtxtext, newtxmath}
\usepackage{braket}
\usepackage{bbold}
\usepackage{bibunits}
\usepackage{hyperref}
\usepackage{xcolor}

\begin{document}

\catchline{}{}{}{}{}

\title{The Role of Localizable Concurrence in Quantum Teleportation Protocols}

\author{Mirko Consiglio}

\address{Department of Physics, University of Malta\\
University Ring Road, Msida MSD2080, Malta\\
mirko.consiglio.16@um.edu.mt}

\author{Louis Zammit Mangion}

\address{Department of Physics, University of Malta\\
University Ring Road, Msida MSD2080, Malta\\
louis.zammit-mangion@um.edu.mt}

\author{Tony John George Apollaro}

\address{Department of Physics, University of Malta\\
University Ring Road, Msida MSD2080, Malta\\
tony.apollaro@um.edu.mt}

\maketitle

\begin{history}
\received{Day Month Year}
\revised{Day Month Year}
\end{history}

\begin{abstract}
Teleporting an unknown qubit state is a paradigmatic quantum information processing task revealing the advantage of quantum communication protocols over their classical counterpart. For a teleportation protocol using a Bell state as quantum channel, the resource has been identified to be the concurrence. However, for mixed multi-partite states the lack of computable entanglement measures has made the identification of the quantum resource responsible for this advantage more challenging. Here, by building on previous results showing that localizable concurrence is the necessary resource for controlled quantum teleportation, we show that any teleportation protocol using an arbitrary multi-partite state, that includes a Bell measurement, requires a non-vanishing localizable concurrence between two of its parties in order to perform better than the classical protocol. By first analyzing the GHZ channel and GHZ measurement teleportation protocol, in the presence of GHZ-symmetric-preserving noise, we compare different multi-partite entanglement measures with the fidelity of teleportation, and we find that the protocol performs better than the classical protocol when all multi-partite entanglement measures vanish, except for the localizable concurrence. Finally, we extend our proof to an arbitrary teleportation protocol with an arbitrary multi-partite entangled channel.
\end{abstract}

\keywords{localizable concurrence; quantum teleportation; GHZ states; quantum circuits.}

\markboth{Consiglio Mirko, Louis Zammit Mangion,  Apollaro Tony John George}
{The Role of Localizable Concurrence in Quantum Teleportation Protocols}

\date{\today}
\maketitle

\section{Introduction}\label{Intro}

Entanglement is the main defining feature of quantum mechanics, allowing quantum  systems to exhibit correlations having no classical counterpart \cite{Plenio_2007}. These correlations constitute a resource that can be exploited to perform several Quantum Information Processing (QIP) tasks, such as quantum computation \cite{Steane_1998}, quantum communication \cite{Gisin_2007} and quantum cryptography \cite{Bennett_2014}. In particular, entanglement is a key resource for quantum teleportation \cite{Bennett_1993, Pirandola2015}, where quantum information is transmitted from one place to another using a classical communication channel and the entanglement shared between the sender and receiver.

Since the inception of quantum teleportation via a Bell state \cite{Bennett_1993}, there has been continuous ongoing research in developing protocols allowing for more complex and elaborate teleportation schemes, such as controlled quantum teleportation, where one or more agents' collaboration is needed in order to complete the teleportation protocol, and quantum error-detection, where the aim is to ascertain whether the initial shared entanglement has been subject to noise during the execution of the protocol, undermining the fidelity of teleportation. Most of these advanced teleportation protocols involve multi-partite entanglement. The perception that some complex teleportation protocols are not feasible, or are resource-inefficient, when relying only on bipartite entanglement, has contributed to the investigation of the vast, and still largely unknown, field of multi-partite entanglement. 

In order to gauge the efficiency of a teleportation protocol, a widely used figure of merit is the fidelity \cite{Jozsa_1994}. In the ideal case, the fidelity of teleportation $F$ attains the maximum of $F = 1$. However, as the faithfulness of teleportation using a classical channel can never exceed $F = \frac{2}{3}$ \cite{Popescu_1994, Massar_1995}, every teleportation protocol achieving a fidelity larger than that classical limit exploits the entanglement between the sender and the receiver as a resource. This is evident in the teleportation protocol with a two-qubit mixed channel, where the fidelity of teleportation is bounded by\cite{Verstraete_2002};
\begin{equation}
\label{eq:fid}
\max\left\{\dfrac{3 + \mathcal{C}}{6}, \dfrac{1 + 2\mathcal{C}}{3} \right\} \leq F \leq \dfrac{2 + \mathcal{C}}{3} \,,
\end{equation}
where $\mathcal{C}$ represents the concurrence, a bipartite entanglement measure for arbitrary two-qubit states \cite{Wootters_1998}. In this scenario, the upper bound is attained by the maximally entangled  Bell states, whereas, in the absence of entanglement one obtains the classical upper bound of $\frac{2}{3}$.

With the theoretical and experimental advances in QIP, great attention has been dedicated to the investigation of entanglement in multi-partite quantum systems \cite{Horodecki2009,Guhne2009} and to their use as a resource for teleportation protocols. The first steps in this direction have involved creating teleportation schemes utilizing tripartite channels, i.e. using the entanglement in a three-qubit state as quantum resource \cite{Karlsson_1998,Lee2005,Agrawal_2006,Yang2009}. Tripartite systems can be represented in three separate ways; completely separable, biseparable, or tripartite entangled states \cite{Dur2000}. In the case of bipartite systems, only the four Bell states, which are equivalent under Local Operations and Classical Communication (LOCC) \cite{Plenio_2007}, are maximally entangled. However, when one ventures into the realm of three or more qubits, one is able to witness different classes of maximally entangled states \cite{Ac_n_2001}. For tripartite systems, we have the family of Greenberg-Horne-Zeilinger (GHZ) states, and the family of W states,
\begin{equation}
\ket{\text{GHZ}} = \dfrac{1}{\sqrt{2}}(\ket{000} + \ket{111})  \,, \ \ket{\text{W}} = \dfrac{1}{\sqrt{3}}(\ket{001} + \ket{010} + \ket{100})
\end{equation}
which are inequivalent under Stochastic LOCC (SLOCC) \cite{Dur2000}, that is under LOCC, but without demanding that the state has to be obtained with certainty. GHZ states have been utilized in tests of local hidden-variable theories \cite{greenberger2007}, and have also been used in superdense coding applications and cryptographic conferencing protocols \cite{Bose1998}. In fact, the family of GHZ states was initially proposed by Greenberger, Horne and Zeilinger to examine non-locality without making use of an inequality, providing a stronger statement of non-locality than Bell's Theorem \cite{QuantumInformation}. One known benefit of using $n$-qubit GHZ states over a Bell state, to teleport qubits, is that they can be used for partial error-detection. Implementing the post-selection after detection of a bit-flip (or phase-flip) error leads to an overall improved fidelity over the standard teleportation scheme involving a Bell channel \cite{Moreno_2018}.

The inequivalence of the two entanglement classes is also witnessed from a resource point of view, as a GHZ state can be utilized for the deterministic teleportation of one qubit, whereas a W state accomplishes the task only with probability $\frac{2}{3}$ \cite{Joo2003}. Moreover, the presence of a third qubit in the GHZ state allows for an increase in the complexity of the teleportation protocol when compared with the standard one using a Bell state. Karlsson and Bourennane \cite{Karlsson_1998} proposed the Controlled Quantum Teleportation (CQT) protocol utilizing a GHZ state. It involves a sender, a receiver and a controller, with quantum teleportation being realized only by the participation of the controller. In the absence of the controller's collaboration in the protocol, the fidelity of CQT is bounded above by the classical limit. Interestingly, in Ref. \citen{Barasinski_2018}, it has been shown that the localizable concurrence \cite{Verstraete_2004, Popp_2005} between two qubits of the GHZ-state is a necessary resource in CQT.

In this paper we build on their result by investigating the necessary resources for the quantum teleportation protocol utilizing a GHZ channel \cite{Eltschka_2014} and a GHZ measurement \cite{Cunha_2019}, while exposing the channel to GHZ-symmetry-preserving decoherence. The typical protocol setup will consist of generating a GHZ state capable of achieving teleportation, followed by applying noise that retains the GHZ symmetry in the quantum channel. We find that the localizable concurrence is a necessary resource for every teleportation protocol which can be decomposed into a quantum circuit including a Bell measurement. Our result encompasses the original CQT protocol \cite{Karlsson_1998, Barasinski_2018}, and in addition applies to teleportation protocols involving an arbitrary number of qubits, and is able to extend CQT to involve an arbitrary number of controllers while still utilizing the same quantum resource.

The paper is organized as follows; In Section \ref{T}, we will review the GHZ channel and measurement teleportation protocol, followed by an overview of the GHZ-symmetric states in Section \ref{GHZsym}, and the relevant entanglement monotones and teleported states in the presence of dephasing and depolarizing noise in Section \ref{Noise}. In Section \ref{locent} we present our main results and discuss the concept of localizable concurrence in the GHZ channel and measurement teleportation protocol, and a comparison is made with CQT and other teleportation protocols. Here we demonstrate that any general multi-partite teleportation scheme involving a Bell measurement as a component of the protocol, necessitates the use of localizable concurrence as a resource in quantum teleportation. Finally, in Section \ref{S.Conc} conclusions are drawn.

\section{GHZ Channel and Measurement Teleportation Protocol} \label{T}

A teleportation protocol is a QIP task able to transfer an unknown quantum state from one place to another using classical communication and pre-established entangled states between the sender and receiver. The first teleportation protocol that has been proposed involves a Bell state shared between two parties, generally named Alice and Bob \cite{Bennett_1993}, aimed at teleporting the unknown state of a qubit. A Bell measurement is performed on the qubit containing the message, and Alice's qubit being part of the Bell state. Then, two bits of classical information containing the measurements' outcome are transmitted to Bob, enabling him to perform a unitary transformation on his qubit to complete the protocol. The GHZ channel and measurement teleportation protocol achieves the same results, however it operates by using both a GHZ channel and a GHZ measurement \cite{Karlsson_1998} as outlined in the following derivation.

The GHZ state is defined as 
\begin{equation}\label{E.GHZ}
\ket{\text{GHZ}} = \dfrac{1}{\sqrt{2}}(\ket{000} + \ket{111}) = \dfrac{1}{\sqrt{2}}\sum\limits_{i=0}^{1}\ket{iii} \,.
\end{equation}

Suppose Alice wants to teleport a single qubit, denoted by the subscript $1$, to Bob, through a GHZ state. In this teleportation protocol Alice possesses two of the GHZ qubits, denoted by the subscripts $2$ and $3$, while Bob has the other GHZ qubit, denoted by the subscript $4$. The general state of qubit 1 can be written as
\begin{equation}
\ket{\psi}_1 = \alpha_0\ket{0}_1 + \alpha_1\ket{1}_1 = \sum\limits_{i=0}^{1}\alpha_i\ket{i}_1 \,.
\label{eq:tpstate}
\end{equation}
Combining Eqs. \eqref{E.GHZ} and \eqref{eq:tpstate}, the state of the entire system reads
\begin{equation}
\ket{\Psi}_{1234} = \dfrac{1}{\sqrt{2}}\sum\limits_{i, j=0}^{1}\alpha_i\ket{ijjj}_{1234} \,.
\end{equation}
Rewriting the above equation by separating the qubits that will be part of the GHZ measurement, we end up with
\begin{equation}
\ket{\Psi}_{1234} = \dfrac{1}{\sqrt{2}}\sum\limits_{i, j=0}^{1}\alpha_i\ket{ijj}_{123} \otimes \ket{j}_4 \,.
\end{equation}
A GHZ measurement can be performed by taking the inner product of
\begin{equation}
\ket{\phi_{lmn}} = \dfrac{1}{\sqrt{2}}\sum\limits_{k=0}^{1}(-1)^{lk}\ket{k,k \oplus m,k \oplus n} \,,
\end{equation}
with the first three qubits of the state $\ket{\Psi}_{1234}$ \cite{Cunha_2019}. This results in Bob acquiring the state 
\begin{equation}
\ket{\eta_{lmn}}_4 = \dfrac{1}{2}\sum\limits_{i, j,k=0}^{1}\alpha_i(-1)^{kl}\braket{k,k \oplus m,k \oplus n|ijj}_{123} \otimes \ket{j}_4 = \sum\limits_{k=0}^{1}\alpha_k(-1)^{kl}\ket{k \oplus n}_4 .
\label{eq:tpmeas}
\end{equation}
Thus, Bob needs to perform unitary operations depending on the measured values of $l, m$ and $n$.

Note that  Alice needs only to send two bits of information since we require that $k \oplus m = j = k \oplus n$, meaning that the measurement results of qubits 2 and 3 must be the same, i.e. $n = m$. Should this not be the case, a bit-flip has occurred on one of those qubits, meaning that the teleported state can be rejected. This means that partial bit-flip errors can be detected in the teleportation protocol, \footnote{It is useful to note that the teleportation protocol can be adjusted to detect phase-flip errors instead of bit-flip errors, which can prove to be advantageous if the frequency of phase-flip errors is higher than the corresponding bit-flip ones} resulting in an improved overall fidelity \cite{Moreno_2018}, which is fundamental for teleportation protocols in ascertaining that the transmitted information is correct. Finally, Bob can retrieve the state in Eq. \eqref{eq:tpstate} by applying the unitary ${\sigma_x}^{n}{\sigma_z}^{l}$ to the state in Eq. \eqref{eq:tpmeas}, resulting in
\begin{equation}
\ket{\psi}_4 = {\sigma_x}^{n}{\sigma_z}^{l}\ket{\eta_{lmn}}_4 = \sum\limits_{k=0}^{1}\alpha_k\ket{k}_4 \,,
\end{equation}
which is exactly the state  as required.

\subsection{GHZ-Symmetric States} \label{GHZsym}

Let us move to a specific family of tripartite high-ranked mixed states, namely the GHZ-symmetric states \cite{Eltschka_2012, Eltschka_2014}. These states are ideal for analyzing the effects of noise on QIP tasks as they have well-defined analytical entanglement monotones, allowing us to derive quantitative relations between the fidelity of the teleportation protocol and different entanglement measures. 

\begin{figure}[ht]
\centering
\includegraphics[width=\textwidth]{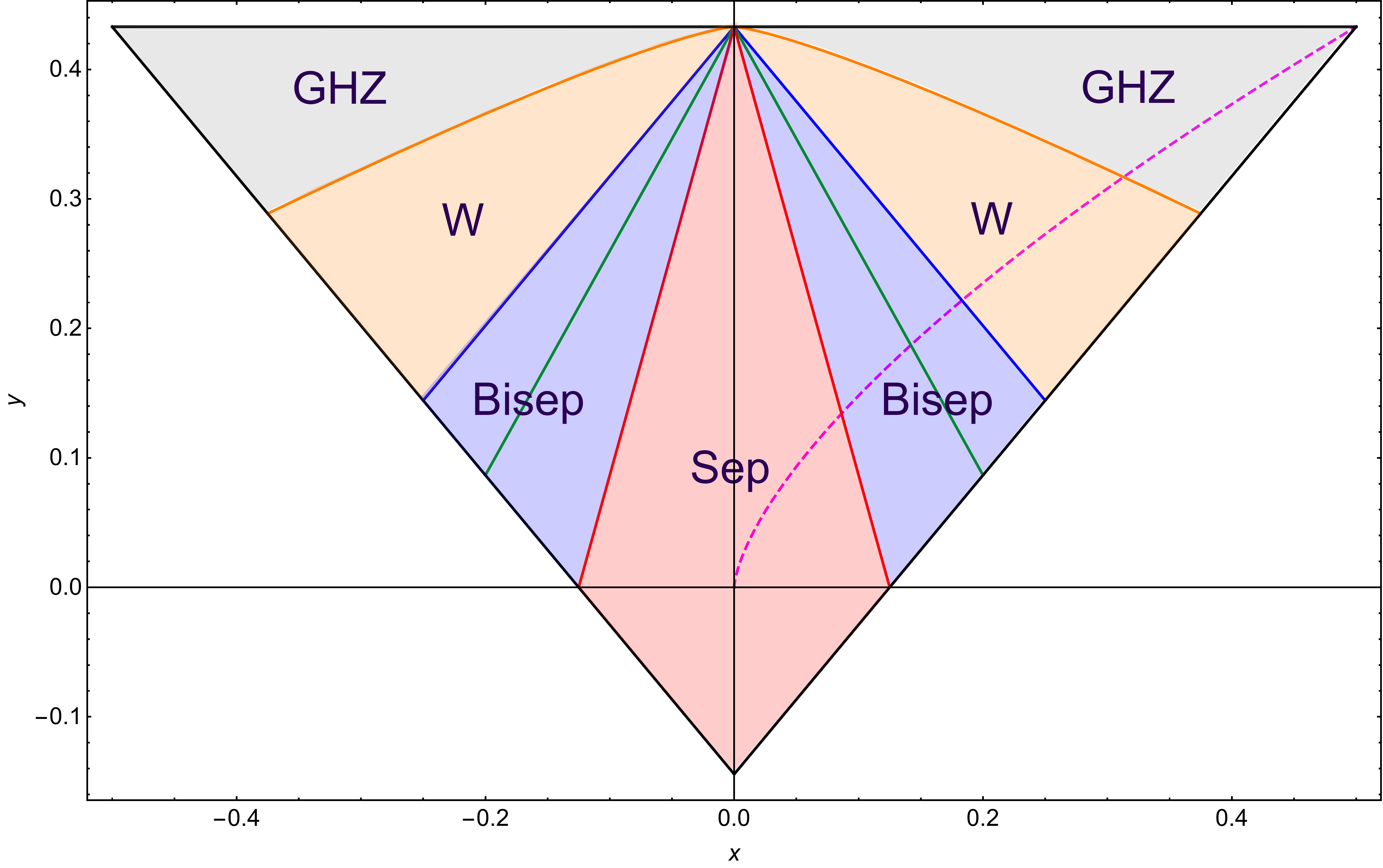}
\caption{Entanglement classes of the GHZ-symmetric states, where on the top two corners lie the $\protect\ket{\text{GHZ}^\pm}$ states. The red kite in the middle represents the separable states. The two blue triangles are the biseparable states. The two yellow triangle-like shapes are the W-class entangled states, bounded by the orange, curved, GHZ-W line. The last two gray triangle-like shapes represent the GHZ-class entangled states. The dashed magenta line represents the path taken by a GHZ state exposed to depolarizing noise as described in Section \ref{depolarizing}. The solid green line separates GHZ-symmetric states which have zero localizable concurrence from those with non-zero value. Figure adapted from Ref. \citen{Eltschka_2014}.}
\label{fig:triangle}
\end{figure}

The GHZ-symmetric family contains all tripartite mixed states which are invariant under the operations listed below;
\begin{enumerate}
\item Qubit permutations.
\item Simultaneous three-qubit-flips.
\item Qubit rotations around the $z$ axis of the form
\begin{equation}
U(\phi_1, \phi_2) = e^{i\phi_1\sigma_z} \otimes e^{i\phi_2\sigma_z} \otimes e^{-i(\phi_1 + \phi_2)\sigma_z} \,.
\end{equation}
\end{enumerate}
The two GHZ-symmetric pure states are  $\ket{\text{GHZ}^+}=\frac{(\ket{000}+ \ket{111})}{\sqrt{2}}$, and the sign-flipped state $\ket{\text{GHZ}^-} \equiv \frac{(\ket{000} - \ket{111})}{\sqrt{2}}$. The complete set of GHZ-symmetric states consists of mixtures of the above two states and the mixed state \linebreak $\rho = \sum_{ijk = 001}^{110} \ket{ijk}\bra{ijk}$. A GHZ-symmetric state $\rho^{GS}$ can be fully described by two real parameters;
\begin{eqnarray}
\label{eq:x}
x\left(\rho^{GS}\right) &=& \dfrac{1}{2}\left( \bra{\text{GHZ}^+}\rho^{GS}\ket{\text{GHZ}^+} - \bra{\text{GHZ}^-}\rho^{GS}\ket{\text{GHZ}^-} \right) \,, \\[1ex]
\label{eq:y}
y\left(\rho^{GS}\right) &=& \dfrac{1}{\sqrt{3}}\left( \bra{\text{GHZ}^+}\rho^{GS}\ket{\text{GHZ}^+} + \bra{\text{GHZ}^-}\rho^{GS}\ket{\text{GHZ}^-} - \dfrac{1}{4} \right) \,.
\end{eqnarray}
A general GHZ-symmetric state can then be written as
\begin{equation}
\begin{split}
\rho^{GS}(x, y) = &\left(\dfrac{2}{\sqrt{3}}y + x \right)\ket{\text{GHZ}^+}\bra{\text{GHZ}^+} + \\ &\left(\dfrac{2}{\sqrt{3}}y - x \right)\ket{\text{GHZ}^-}\bra{\text{GHZ}^-}  + \dfrac{\sqrt{3} - 4y}{8\sqrt{3}}\mathbb{1}_8 \,,
\label{eq:GHZsym}
\end{split}
\end{equation}
where $\mathbb{1}_8$ is the $8 \times 8$ identity matrix. This parameterization constructs Fig. \ref{fig:triangle}, where the completely mixed state is located at the origin.
\text{}

Any point in this triangle describes a GHZ-symmetric state, with each region describing a different entanglement class with respect to SLOCC. Particularly, the GHZ-class states are bounded below by the parameterized line;
\begin{equation}
x^W = \dfrac{v^5+8v^3}{8(4-v^2)}, \
y^W = \dfrac{\sqrt{3}}{4}\dfrac{4-v^2-v^4}{4-v^2} \,,
\label{GHZW}
\end{equation}
where $v \in [-1, 1]$. Given the coordinates specified in Eqs. \eqref{eq:x} and \eqref{eq:y}, one is able to determine the three-tangle, an entanglement monotone that measures genuine tripartite entanglement of the GHZ-class states \cite{Coffman_2000} as follows; one first determines the straight line that connects the GHZ$^+$ state at $(\frac{1}{2}, \frac{\sqrt{3}}{4})$ with the point $(x, y)$. This line intersects the GHZ-W line at the point $(x^W, y^W)$, from which the three-tangle $\tau_3\left(\rho^{GS}(x, y)\right)$ is determined by
\begin{equation}
\tau_3(x, y) = 
\begin{cases}
0 \,, & \text{for } x < x^W\,, y < y^W \,, \\
\dfrac{x - x^W}{\frac{1}{2} - x^W} \,, & \text{otherwise.}
\end{cases}
\label{eq:3tangle}
\end{equation}
Below this line the GHZ-symmetric state turns into a W-class state. The curve described by Eq. \eqref{GHZW} is in fact typically called the GHZ-W line. The lower bound of the W-class states is characterized by the vanishing of the Genuine Multi-partite Entanglement (GME) concurrence \cite{D_r_1999, de_Vicente_2011, Huber_2014}, which is given as
\begin{equation}
\mathcal{C}_{\text{GME}}\left(\rho^{GS}\right) = \max\left\{0, 2|x| + \sqrt{3}y - \dfrac{3}{4}\right\}.
\label{eq:GME}
\end{equation}
This line separates the W-class states from the Biseparable-class states, which is denoted as the W-B line. Finally, the lower bound for Biseparable-class states is characterized by the vanishing of the negativity \cite{Horodecki_1998}, given by
\begin{equation}
\mathcal{N}\left(\rho^{GS}\right) = \max\left\{0, |x| + \dfrac{1}{2\sqrt{3}}y - \dfrac{1}{8}\right\}.
\label{eq:Neg}
\end{equation}
Below this line lie the separable-class states.

To gauge the efficiency of the teleportation protocol in the presence of noise, we employ the fidelity \cite{Jozsa_1994} between the initial state Alice wants to teleport, $\ket{\psi} = \cos\left(\frac{\vartheta}{2}\right)\ket{0} + e^{i\varphi}\sin\left(\frac{\vartheta}{2}\right)\ket{1}$, and Bob's state after the teleportation protocol is completed, $\rho_T$.
\begin{equation}
F(\ket{\psi}, \rho_T) = \bra{\psi}\rho_T\ket{\psi} .
\end{equation}
When using a GHZ-symmetric state channel for the teleportation protocol, the fidelity reads
\begin{equation}
F(\vartheta) = \dfrac{1}{6} \left(3 + 3 |x| + 2 \sqrt{3} y + (2 \sqrt{3} y - 3 |x|) \cos\left(2\vartheta\right)\right) \,.
\end{equation}
Notice that the state dependence of the fidelity can be removed by considering the average fidelity of teleportation instead:
\begin{equation}
\bar{F} = \dfrac{1}{18} \left(9 + 12 |x| + 4 \sqrt{3} y\right) \,,
\end{equation}
resulting in a fidelity dependent only on the GHZ-symmetric quantum channel used in the protocol.

\subsection{Fidelity of teleportation with noisy channels}\label{Noise}

The fidelity of teleportation with noisy tripartite channels has been extensively investigated. \cite{Jung2008,Li2010a,Chun_2010,Hu2011,Liang2015}
Here we will restrict our attention to GHZ-symmetry-preserving noise acting on each qubit of the GHZ state before the teleportation protocol is applied. The effect of both dephasing, and depolarizing noise, on the entanglement measures and the fidelity of the teleportation protocol will be analyzed.

\subsubsection{Dephasing noise}

The first type of noise to be discussed is dephasing noise. The Kraus operators of a phase damping channel are the following;
\begin{equation}
K_0 = \begin{pmatrix}
1 & 0 \\
0 & \sqrt{1 - p}
\end{pmatrix} \,, \ K_1 = \begin{pmatrix}
0 & 0 \\
0 & \sqrt{p}
\end{pmatrix}
\end{equation}
where $p \in [0, 1]$ \cite{NC2011}. Thus, homogeneous phase damping acting on each qubit of a GHZ state results in the following mixed state:
\begin{equation}
\rho = \dfrac{1}{2}\left(
\begin{array}{cccccccc}
1 & 0 & 0 & 0 & 0 & 0 & 0 & (1-p)^{\frac{3}{2}} \\
0 & 0 & 0 & 0 & 0 & 0 & 0 & 0 \\
0 & 0 & 0 & 0 & 0 & 0 & 0 & 0 \\
0 & 0 & 0 & 0 & 0 & 0 & 0 & 0 \\
0 & 0 & 0 & 0 & 0 & 0 & 0 & 0 \\
0 & 0 & 0 & 0 & 0 & 0 & 0 & 0 \\
0 & 0 & 0 & 0 & 0 & 0 & 0 & 0 \\
(1-p)^{\frac{3}{2}} & 0 & 0 & 0 & 0 & 0 & 0 & 1
\end{array}
\right) \,.
\end{equation}\\
Using equations \eqref{eq:x} and \eqref{eq:y}, for the above state we get the parameters
\begin{equation}
x(\rho) = \dfrac{1}{2}(1-p)^\frac{3}{2} \,, \ y(\rho) = \dfrac{\sqrt{3}}{4} \,.
\end{equation}
Now, using the procedure described in Section \ref{GHZsym}, we find that every point described with the above equation lies on a horizontal line between the GHZ$^+$ and GHZ$^-$ points, intersecting with $(x^W, y^W) = (0, \frac{\sqrt{3}}{4})$. Applying Eq. \eqref{eq:3tangle}, we find that the three-tangle results
\begin{equation}
\label{eq:dephasing_tau}
\tau_3(\rho) = (1 - p)^\frac{3}{2} \,.
\end{equation}
Similarly using Eqs. \eqref{eq:GME} and \eqref{eq:Neg} to find the GME concurrence and negativity, respectively, we get
\begin{eqnarray}
\label{eq:dephasing_GME}
\mathcal{C}_\text{GME}(\rho) &=& (1 - p)^\frac{3}{2} \,, \\
\label{eq:dephasing_N}
\mathcal{N}(\rho) &=& \frac{1}{2}(1 - p)^\frac{3}{2} \,.
\end{eqnarray}
The teleported state $\rho_T$, utilizing a dephased GHZ channel, becomes
\begin{equation}
\rho_T = \left(
\begin{array}{cc}
\cos ^2\left(\frac{\vartheta }{2}\right) & \frac{1}{2} (1-p)^\frac{3}{2} e^{i \varphi } \sin (\vartheta ) \\[1ex]
\frac{1}{2} (1-p)^\frac{3}{2} e^{-i \varphi } \sin (\vartheta ) & \sin ^2\left(\frac{\vartheta }{2}\right)
\end{array}
\right) \,,
\end{equation}
with an average fidelity of
\begin{equation}
\bar{F} = \dfrac{1}{3} \left(2 + (1-p)^\frac{3}{2}\right) \,.
\end{equation}
Thus, when $p$ goes to unity, $\bar{F} = \frac{3}{2}$, which is exactly the average fidelity achieved by using only a classical communication channel to transmit information.

Clearly the dephasing channel does not allow for discerning the quantum resource in use for attaining quantum teleportation, since all three entanglement measures, given in Eqs. \eqref{eq:dephasing_tau}, \eqref{eq:dephasing_GME} and \eqref{eq:dephasing_N}, vanish simultaneously at $p = 1$. As such we require investigating another type of noise capable of discriminating among the quantum resources in the teleportation protocols via a GHZ channel.

\subsubsection{Depolarizing noise} \label{depolarizing}

\begin{figure}[ht]
\centering
\includegraphics[width=0.49\textwidth]{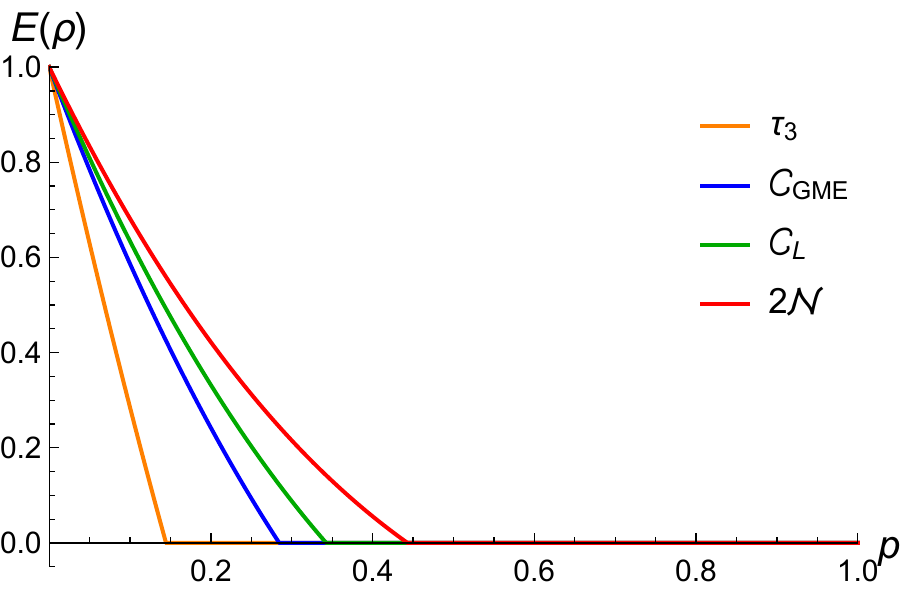}
\includegraphics[width=0.49\textwidth]{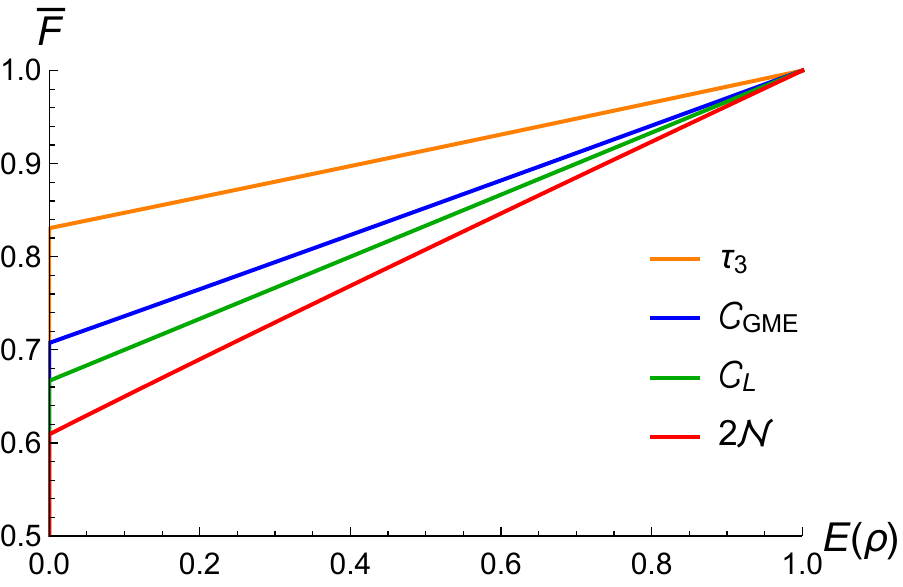}
\caption{(left) Entanglement measures as a function of the rate of depolarization, $p$. (right) Teleportation fidelity \eqref{eq:depolfid} as a function of different entanglement measures. One can see that the entanglement measures decay to zero in the order determined by the SLOCC classes the depolarizing GHZ state passes through. Looking at the right panel, a tripartite channel can have GME concurrence and three-tangle zero, yet the teleported state has a fidelity greater than $\frac{2}{3}$. The classical bound is only reached when the localizable concurrence is equal to zero.}
\label{fig:entmeas}
\end{figure}

We turn our focus to a GHZ state under depolarizing noise. The Kraus operators of a depolarizing channel are
\begin{equation}
\begin{split}
K_0 &= \dfrac{1}{\sqrt{2}}\left(\begin{array}{lc}
1 & 0 \\
0 & \sqrt{1 - p}
\end{array} \right) \,, \ K_1 = \dfrac{1}{\sqrt{2}}\left(\begin{array}{lc}
0 & \sqrt{p} \\
0 & 0
\end{array} \right) \,, \\[1ex]
K_2 &= \dfrac{1}{\sqrt{2}}\left(\begin{array}{cl}
\sqrt{1 - p} & 0 \\
0 & 1
\end{array} \right) \,, \ K_3 = \dfrac{1}{\sqrt{2}}\left(\begin{array}{cl}
0 & 0 \\
\sqrt{p} & 0
\end{array} \right) \,,
\end{split}
\end{equation}
where $p \in [0, 1]$ \cite{NC2011}. The noisy GHZ-symmetric state reads
\begin{equation}
\rho = \left(
\renewcommand*{\arraystretch}{0.9}
\begin{array}{cccccccc}
a & 0 & 0 & 0 & 0 & 0 & 0 & c \\
0 & b & 0 & 0 & 0 & 0 & 0 & 0 \\
0 & 0 & b & 0 & 0 & 0 & 0 & 0 \\
0 & 0 & 0 & b & 0 & 0 & 0 & 0 \\
0 & 0 & 0 & 0 & b & 0 & 0 & 0 \\
0 & 0 & 0 & 0 & 0 & b & 0 & 0 \\
0 & 0 & 0 & 0 & 0 & 0 & b & 0 \\
c & 0 & 0 & 0 & 0 & 0 & 0 & a
\end{array}
\right) \,,
\end{equation}
where $a = \frac{(4-3(2-p)p)}{8}$, $b = \frac{(2-p)p}{8}$, $c = \frac{(1-p)^3}{2}$. The mixed state traces the dashed magenta path in Fig. \ref{fig:triangle} as $p$ goes from zero to unity. Once again, using equations \eqref{eq:x} and \eqref{eq:y}, for the above state we get
\begin{equation}
x(\rho) = \dfrac{1}{2}(1-p)^3 \,, \ y(\rho) = \dfrac{\sqrt{3}}{4}(1-p)^2 \,.
\end{equation}
Here we require the calculation of the three-tangle numerically, since the resulting equations require the solution of a quintic polynomial. In addition, for an analytical comparison, an accurate formula for the three-tangle derived from the analytical lower bound in Ref. \citen{Eltschka_2014}, is given here;
\begin{equation}
\label{eq:approx3tangle}
\tilde{\tau}_3(\rho) = \max\left\{0, \dfrac{1}{7}(7 - 54p + 39p^2 - 8p^3)\right\} \,.
\end{equation}
Note that the difference between the exact numerical value and the lower bound is less than $2 \times 10^{-3}$, and when the lower bound is zero, the exact three-tangle is also zero. Now, using Eqs. \eqref{eq:GME} and \eqref{eq:Neg} to find the GME concurrence and negativity, respectively, we get
\begin{eqnarray}
\label{eq:gmeconc}    
\mathcal{C}_\text{GME}(\rho) &=& \max\left\{0, \dfrac{1}{4}(4 - 18p + 15p^2 - 4p^3) \right\} \,, \\[1ex]
\label{eq:neg2}
\mathcal{N}(\rho) &=& \max\left\{0, \dfrac{1}{8}(4 - 14p + 13p^2 - 4p^3)\right\} \,.
\end{eqnarray}
Fig. \ref{fig:entmeas} (left panel) shows the plots of these three entanglement measures along with the localizable concurrence, which shall be derived in the following section. Given a depolarized GHZ channel, the teleported state $\rho_T$ becomes
\begin{equation}
\rho_T = \dfrac{1}{2}\left(
\begin{array}{cc}
1+(1-p)^2 \cos (\vartheta ) & (1-p)^3 e^{i \varphi } \sin (\vartheta )
\\[1ex]
(1-p)^3 e^{-i \varphi } \sin (\vartheta ) & 1-(1-p)^2 \cos (\vartheta )
\\
\end{array}
\right) \,,
\end{equation}
with an average fidelity of
\begin{equation}
\label{eq:depolfid}
\bar{F} = \dfrac{1}{6}\left(6 - 8 p + 7 p^2 - 2 p^3\right) \,.
\end{equation}
Here we find that $\bar{F} = \frac{1}{2}$ as $p$ goes to unity.

Letting $G, W, B$ and $S$ denote the sets of GHZ-, W-, biseparable- and separable-class states, we find that the three-tangle \eqref{eq:approx3tangle} equals zero when $p_{G\backslash W} \approx 0.144204$, which achieves an average fidelity of $\bar{F}_{G\backslash W} \approx 0.830989$, which is larger than $\frac{2}{3}$. The same procedure can be applied for the GME concurrence \eqref{eq:gmeconc} and the negativity \eqref{eq:neg2}. The GME concurrence equals zero when $p_{W\backslash B} \approx 0.284595$, achieving an average fidelity of $\bar{F}_{W\backslash B} \approx 0.707350$, while the negativity equals zero when $p_{B\backslash S} \approx 0.443307$, with an average fidelity of $\bar{F}_{B\backslash S} \approx 0.609159$. Only when $p_Q \approx 0.342702$ does $\bar{F} = \frac{2}{3}$.

Fig. \ref{fig:entmeas} (right panel) shows the teleportation fidelity as a function of different entanglement measures. As one can see, the maximum classical fidelity is reached when the depolarized GHZ state is in the biseparable-class, yet not all biseparable mixed states are able to achieve a fidelity higher than $\frac{2}{3}$. This implies that GME concurrence, and consequently three-tangle, are not necessary for the GHZ channel and measurement teleportation protocol. The quantum resource necessary to outperform classical communication protocols turns out to be the localizable concurrence, which shall be discussed in the following section.

\section{Localizable Concurrence} \label{locent}

It has been already established that localizable concurrence is a necessary resource for the CQT protocol \cite{Barasinski_2018}. Localizable concurrence is concretely defined in Section IIA of Ref. \citen{Popp_2005}, by considering the concurrence as the entanglement measure applied to the definition. The localizable concurrence of a quantum channel can then be roughly defined as the maximum possible concurrence generated between two qubits, specifically the sender and the receiver, given that optimal local measurements are performed on all other qubits present in the quantum channel. Here we prove that the localizable concurrence is also the necessary resource for the quantum teleportation protocol using a GHZ state and GHZ measurement. 

\begin{figure}[ht]
\centering
\includegraphics[width=0.75\textwidth]{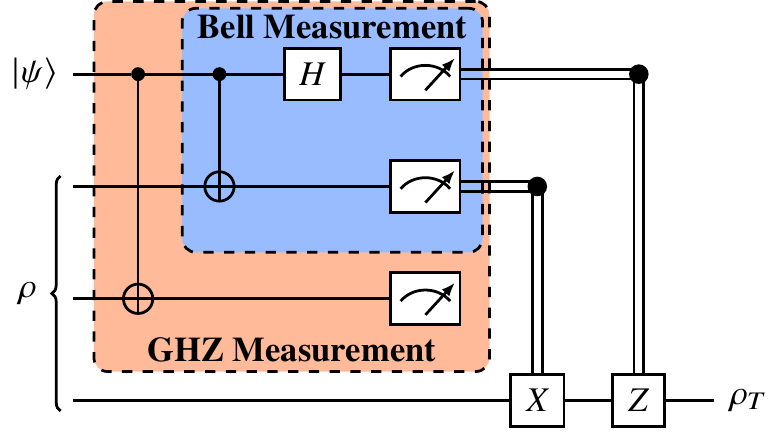}
\caption{Quantum circuit for teleporting a qubit containing the state $\ket{\psi}$, using a tripartite state $\rho$. The top three lines of the circuit represent Alice's message and tripartite qubits. The bottom line is Bob's tripartite qubit.}
\label{fig:teleport_GHZ}
\end{figure}

We shall consider the teleportation protocol from the perspective of a quantum circuit. The quantum circuit for the tripartite channel teleportation protocol with GHZ measurement is given in Fig. \ref{fig:teleport_GHZ}, and we highlight the decomposition of the GHZ measurement into a Bell measurement and a preceding $CX$ gate. As in the standard teleportation scheme \cite{Bennett_1993} concurrence is the necessary quantum resource and a Bell measurement is an essential part of the protocol, identifying in a teleportation scheme via multi-partite entanglement a multi-qubit measurement decomposition involving Bell measurements paves the way to the identification of the necessary quantum resource.

This measurement decomposition constitutes the key argument of our proof as it reduces the GHZ channel and measurement protocol to an effective standard teleportation protocol where the localizable concurrence is the necessary quantum resource utilized for teleportation. Note that the last measurement also serves to ascertain that both of the measurements on Alice's tripartite state qubits give the same measurement value, that is both ones or zeros. If they differ, then one can be certain that a bit-flip error has occurred in one of the qubits \cite{Moreno_2018}, allowing for the post-selection of noisy measurements, which cannot be carried out by using a Bell channel. This partial error detection leads to a better overall fidelity of teleportation.

Localizable concurrence is not a quantity that is straightforward to evaluate, however following on the lines of Ref. \citen{Barasinski_2018}, we can calculate the average localized concurrence between qubits 2 and 4, being the sender qubit and receiver qubit, respectively. By applying the $CX$ gate between qubits 1 and 3, and then applying an optimal one-qubit orthogonal measurement on qubit 3 (as if qubit 3 is playing the role of the controller within the GHZ measurement circuit decomposition given in Fig. \ref{fig:teleport_GHZ}), the localizable concurrence between the sender and receiver qubits is calculated as follows:
\begin{equation}
\mathcal{C}_L(\rho) = \max_{U}\sum\limits_{t=0}^{1}\bra{t}U \rho_3 U^{\dagger}\ket{t}\mathcal{C}\left(\rho_{24}^{t}\right) \,,
\label{eq:locconc}
\end{equation}
where $\rho' = \ket{\psi}\bra{\psi} \otimes \rho$ represents the state of the message qubit and tripartite channel, and $\rho_{24}^{t}$ represents the state of qubits 2 and 4 after applying a $CX$ gate between qubits 1 and 3, and obtaining the measurement outcome $t$ on qubit 3;
\begin{equation}
\rho_{24}^{t} = \dfrac{\text{Tr}_{3}\left\{ \left(U^{\dagger}\ket{t}\bra{t}U\right)_3\text{Tr}_1\left\{CX_{13}\rho'CX_{13}\right\}\left( U^{\dagger}\ket{t}\bra{t}U\right)_3 \right\}}{\bra{t}U\rho_3 U^{\dagger}\ket{t}} \,,
\label{eq:localdens}
\end{equation}
where the subscripts on the operators denote the qubit on which it is acting. $\bra{t}U\rho_3 U^{\dagger}\ket{t}$ represents the probability of achieving the outcome $t$ on qubit 3 after applying the $CX$ gate, and $U$ is a $2 \times 2$ unitary matrix.

Taking the general form of a $2 \times 2$ unitary matrix (ignoring global phase),
\begin{equation}
\label{eq:U}
U = \left( \begin{array}{cc}
\cos\left(\frac{\theta}{2}\right) & -e^{i\lambda}\sin\left(\frac{\theta}{2}\right) \\[1ex]
e^{i\phi}\sin\left(\frac{\theta}{2}\right) & e^{i(\phi + \lambda)}\cos\left(\frac{\theta}{2}\right)
\end{array} \right) \,,
\end{equation}
where $\theta \in [0, 4\pi], \phi, \lambda \in [0, 2\pi]$, so that we are able to maximize Eq. \eqref{eq:locconc}, we find that for a GHZ-symmetric state $\rho^{GS}$,
\begin{equation}
\rho_{24}^{t} = \left(
\renewcommand*{\arraystretch}{0.8}
\begin{array}{cccc}
a + b & 0 & 0 & c \\[1ex]
0 & d & 0 & 0 \\[1ex]
0 & 0 & d & 0 \\[1ex]
c^* & 0 & 0 & a - b
\end{array}
\right) \,,
\end{equation}
where $a = \frac{1}{4} + \frac{y}{\sqrt{3}}$, $b=(-1)^t\frac{2y\cos(\vartheta)\cos(\theta)}{\sqrt{3}}$, $c = -(-1)^t x \sin (\theta ) (\cos (\lambda )-i \cos (\vartheta) \sin (\lambda ))$, and $d = \frac{1}{4} - \frac{y}{\sqrt{3}}$,
with $\bra{t}U\rho_3U^{\dagger}\ket{t} = \frac{1}{2}$ since $\rho_3 = \frac{1}{2}\mathbb{1}_2$ and $t = 0, 1$. From this we can calculate the average bipartite concurrence between qubits 2 and 4, which is maximized when $\theta = \frac{\pi}{2}$ and $\lambda=0, \pi$:
\begin{equation}
\mathcal{C}_L\left(\rho^{GS}\right) = 2\max\left\{0, |x| + \dfrac{y}{\sqrt{3}} - \dfrac{1}{4}\right\} \,.
\end{equation}
This is exactly the result found in Ref. \citen{Barasinski_2018}. For a depolarized GHZ state $\rho$,
\begin{equation}
\mathcal{C}_L\left(\rho\right) = \max\left\{0, \dfrac{1}{4}\left(2 - 8 p + 7 p^2 - 2 p^3\right)\right\} \,,
\end{equation}
where $\mathcal{C}_L = 0$ when $p = p_Q$ for the depolarized GHZ channel, implying that localizable concurrence between qubits 2 and 4 is a necessary resource for the tripartite channel teleportation protocol utilizing a GHZ measurement. Clearly, because of the permutation symmetry of the GHZ-symmetric states, the role of the qubits is interchangeable.

\begin{figure}[ht]
\centering
\includegraphics[width=0.6\textwidth]{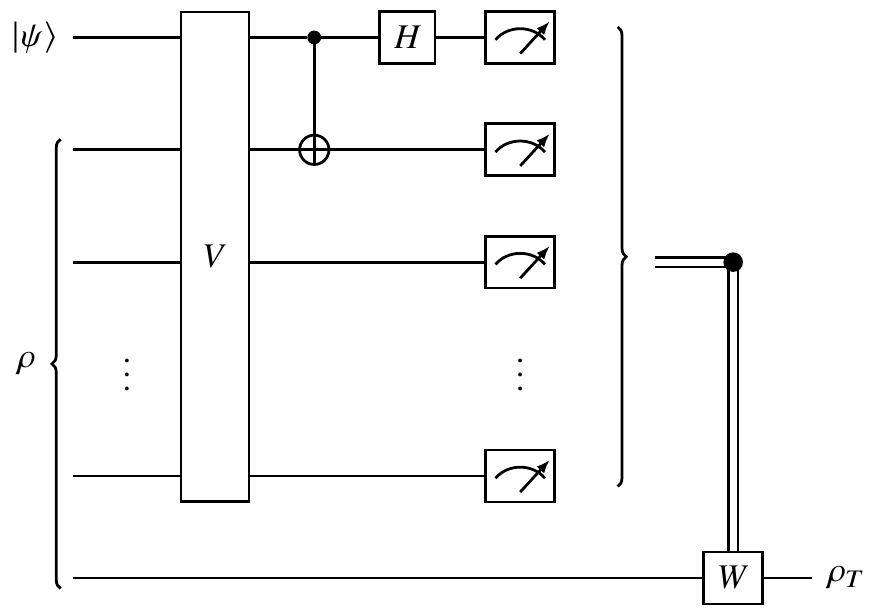}
\caption{Teleportation scheme following the definition for Prop. \ref{thm:prop1}.}
\label{fig:prop1}
\end{figure}

This can be generalized to $n$-partite teleportation protocols of a qubit from a sender to a receiver. Below we provide a proof that the localizable concurrence is a necessary resource for any teleportation scheme requiring a Bell measurement within the set of measurements carried out during the protocol. A quantum circuit scheme, similar to Fig. \ref{fig:teleport_GHZ} for the GHZ measurement protocol, is given in Fig. \ref{fig:prop1}. We generalize Proposition 1 of Ref. \citen{Barasinski_2018} using the same line of reasoning, in order to go beyond the applications of localizable concurrence in CQT protocols only. The main difference is that our proof includes the possibility of applying multi-qubit gates between the message state and the $n$-partite entangled state, and in between the entangled state itself, before any measurements are performed.

\newtheorem{prop}{Proposition}
\begin{prop}
\label{thm:prop1}
Suppose we have a teleportation protocol that can be constructed such as in Fig. \ref{fig:prop1}. The protocol utilizes an entangled $n$-partite state $\rho$ capable of achieving the teleportation of one qubit. Without loss of generality, the initial state $\ket{\psi}$ is stored in qubit 1, with the sender being qubit 2 and the intended receiver being qubit $n + 1$, acquiring the state $\rho_T$ after teleportation. Qubits 3 to $n$ can be under the possession of any combination of the sender and any number of controllers. $V$ contains all the (non-)local operators that are necessary for teleportation, while $W$ is the classically-conditioned operator required to complete the protocol, which depends on the measurement results from the other qubits. Then the fidelity of the teleportation protocol using the n-partite state $\rho$ is
\begin{equation}
\max\left\{\dfrac{3 + \mathcal{C}_L(\rho)}{6}, \dfrac{1 + 2\mathcal{C}_L(\rho)}{3} \right\} \leq F_T(\rho) \leq \dfrac{2 + \mathcal{C}_L(\rho)}{3} \,,
\end{equation}
where $\mathcal{C}_L(\rho)$ is the localizable concurrence of $\rho$ and $F_T(\rho) \equiv F(\ket{\psi}, \rho_T)$.
\end{prop}

\begin{proof}
Define $X$ to be the set of qubits which shall be measured, after applying the necessary set of gates $V$ required in the protocol to induce localized entanglement between the sender $S$ and receiver $R$, i.e. in this case qubits $S = 2$, $R = n + 1$ and $X = \{3, \dots, n\}$. Define $\rho' = \ket{\psi}\bra{\psi} \otimes \rho$ and $\rho_X = \text{Tr}_{[n] \backslash X}\left\{ V_{[n]}\rho'V^\dagger_{[n]}\right\}$, where we define $[n] \equiv \{1, \dots, n\}$. Then
\begin{equation}
\rho_{SR}^t = \dfrac{\text{Tr}_X\left\{\left(U^\dagger\ket{t}\bra{t}U\right)_X \text{Tr}_1 \left\{ V_{[n]}\rho'V^\dagger_{[n]} \right\} \left(U^\dagger\ket{t}\bra{t}U\right)_X\right\}}{\bra{t}U\rho_X U^\dagger\ket{t}} \,,
\end{equation}
where the subscripts on the operators denote the qubit on which it is acting. $U$ is the unitary matrix to be maximized, such that
\begin{equation}
\label{eq:cl}
\mathcal{C}_L(\rho) = \max_{U}\sum\limits_{t}\bra{t}U \rho_X U^{\dagger}\ket{t}\mathcal{C}\left(\rho_{SR}^{t}\right) \,,
\end{equation}
\begin{equation}
F_T(\rho) = \max_{U}\sum\limits_{t}\bra{t}U \rho_X U^{\dagger}\ket{t}F\left(\rho_{SR}^{t}\right) \,.
\end{equation}

Let us first prove the upper bound. Suppose there exists $\tilde{U}$ which maximizes Eq. \eqref{eq:cl}, such that $\tilde{\rho}_{SR}^{t}$ is an optimal two-qubit state. Now from Eq. \eqref{eq:fid},
\begin{equation}
F_T\left(\rho_{SR}^{t}\right) \leq \dfrac{2 + \mathcal{C}\left(\rho_{SR}^{t}\right)}{3} \,,
\end{equation}
where $\mathcal{C}\left(\rho_{SR}^{t}\right)$ is the concurrence of the localized state $\rho_{SR}^{t}$. Thus
\begin{eqnarray}
F_T(\rho) &=& \sum\limits_t \bra{t}\tilde{U}\rho_X \tilde{U}^\dagger\ket{t}F\left(\tilde{\rho}_{SR}^{t}\right) \nonumber \\
&\leq& \sum\limits_t \bra{t}\tilde{U}\rho_X \tilde{U}^\dagger\ket{t}\dfrac{2 +        \mathcal{C}\left(\tilde{\rho}_{SR}^{t}\right)}{3} \nonumber \\
&\leq& \max_U \sum\limits_t \bra{t}U\rho_X U^\dagger\ket{t}\dfrac{2 + \mathcal{C}\left(\rho_{SR}^{t}\right)}{3} \nonumber \\
&=& \dfrac{2 + \mathcal{C}_L(\rho)}{3}\,.
\end{eqnarray}
Now for the lower bound. Suppose that there also exists $\tilde{U}$ which maximizes Eq. \eqref{eq:cl}, then from Eq. \eqref{eq:fid}
\begin{equation}
\max\left\{\dfrac{3 + \mathcal{C}\left(\tilde{\rho}_{SR}^{t}\right)}{6}, \dfrac{1 + 2\mathcal{C}\left(\tilde{\rho}_{SR}^{t}\right)}{3} \right\} \leq F\left(\tilde{\rho}_{SR}^{t}\right) \,.
\end{equation}
Therefore,
	\begin{eqnarray}
    \hspace{2.8cm} & & \hspace{-2.8cm} \max\left\{\dfrac{3 + \mathcal{C}_L(\rho)}{6}, \dfrac{1 + 2\mathcal{C}_L(\rho)}{3} \right\} = \nonumber \\ &=&  \max_{U}\sum\limits_{t}\bra{t}U \rho_X U^{\dagger}\ket{t}\max\left\{\dfrac{3 + \mathcal{C}\left(\rho_{SR}^{t}\right)}{6}, \dfrac{1 + 2\mathcal{C}\left(\rho_{SR}^{t}\right)}{3} \right\} \nonumber \\
	&=& \sum\limits_{t} \bra{t}\tilde{U}\rho_X\tilde{U}^\dagger\ket{t}\max\left\{\dfrac{3 + \mathcal{C}\left(\tilde{\rho}_{SR}^{t}\right)}{6}, \dfrac{1 + 2\mathcal{C}\left(\tilde{\rho}_{SR}^{t}\right)}{3} \right\} \nonumber \\
	&\leq& \sum\limits_{t} \bra{t}\tilde{U}\rho_X\tilde{U}^\dagger\ket{t} F\left(\tilde{\rho}_{SR}^{t}\right) \nonumber \\
	&\leq& \max_{U} \sum\limits_{t}\bra{t}U\rho_X U^\dagger\ket{t}F\left(\rho_{SR}^{t}\right) \nonumber \\
	&=& F_T(\rho)
	\end{eqnarray}
\end{proof}
This scheme encompasses the GHZ channel and measurement teleportation, CQT, and W-class teleportation protocols \cite{Agrawal_2006, Li_2007}. Note that for the standard teleportation scheme utilizing a Bell state and Bell measurement, which is also contained within the proposition above, the concurrence replaces the localizable concurrence. An important remark is that while the unitary matrix in Eq. \eqref{eq:U} allows for the determination of the localizable concurrence, for general $n$-partite states, maximizing Eq. \eqref{eq:cl} over all possible measurements (by means of a general $(n-2)$-qubit unitary matrix) is not a trivial task. At the same time, from Fig. \ref{fig:triangle} it is evident that neither W- nor GHZ-entanglement is necessary for obtaining localizable concurrence, whereas biseparability seems to be a necessary although not sufficient condition. The same appears to hold for $n>3$ qubit states \cite{Barasinski_2018}, although the boundary for non-zero localizable concurrence within the biseparable class of states is unknown.

We already mentioned some advantages of using a GHZ channel over a Bell channel to teleport a single qubit, such as the capability to carry out partial error detection \cite{Moreno_2018} and CQT \cite{Karlsson_1998}. Additionally there is the secret sharing protocol \cite{Hillery1999}, consisting of splitting quantum information between two or more party members, quantum state sharing \cite{Man2007}, controlled dense coding \cite{Hao2001}, synchronization of clocks \cite{Komar2014}, and many others. This implies that there are certain instances where physically preparing \cite{Roos2004, DeLimaBernardo2019} and utilizing a GHZ state is more practical, given their extended applications over Bell states. Moreover, analysis of noisy GHZ channels \cite{Miao2010, Yang2019} demonstrates that theoretical and experimental improvements on QIP protocols increasingly allows the implementation of new technologies that limit the effects of noise on GHZ states. Most of these considerations can be extended to $n$-partite GHZ states, as well as other multi-partite states designed to work with other QIP protocols.

A particularly interesting protocol to consider is the teleportation of a Bell state utilizing a GHZ channel and measurement \cite{Gorbachev2000}. Alice possesses one qubit of the GHZ channel and the Bell state to be teleported, while Bob possesses the other two qubits from the GHZ state. The GHZ measurement is performed over Alice's qubits. Alice then sends two bits of classical information to Bob prompting him to perform local operators on his own two qubits to complete the teleportation protocol. Now assuming that the channel is exposed to GHZ-symmetry-preserving noise, one can evaluate that the fidelity of teleportation is equal to
\begin{equation}
F_T\left(\rho^{GS}\right) = \dfrac{(3 + 12|x| + 4\sqrt{3}y)}{12} \,,
\end{equation}
such that the concurrence of the teleported Bell state is equal to unity if the fidelity is also equal to unity, and the concurrence is equal to zero if the fidelity is equal to $\frac{1}{2}$. Moreover, one can also find that the concurrence of the teleported Bell state is equal to the localizable concurrence of the quantum channel. In the case of GHZ-symmetric channels, by referring to the solid green line in Fig. \ref{fig:triangle}, the concurrence of the teleported Bell state goes to zero at the points where the localizable concurrence of the GHZ-symmetric state vanishes. Note that this teleportation protocol would require two Bell channels needing four qubits in total, as opposed to utilizing one GHZ channel requiring the use of only three qubits, thus making this protocol resource efficient.

\section{Conclusion}\label{S.Conc}

In this paper we have investigated the necessary quantum resource in order to achieve a better-than-classical teleportation fidelity when using a GHZ channel and GHZ measurement protocol. In line with the results of Ref. \citen{Barasinski_2018}, where it has been shown that the localizable concurrence is the necessary resource for the controlled quantum teleportation protocol, we found that for the teleportation protocol we investigated, the localizable concurrence is also the necessary resource required by the protocol. By further analyzing the role of localizable concurrence in quantum teleportation protocols, we have first reviewed and outlined the effects of dephasing and depolarizing noise in GHZ channels. The GHZ-symmetric mixed states obtained through the effect of the considered noise have well-defined entanglement measures, which were utilized in relating quantitatively the fidelity of teleportation and different entanglement measures. We have found that better-than-classical fidelity of teleportation can be achieved whenever the localizable concurrence between two qubits, with one being the sender and the other the receiver, of the GHZ-symmetric state is non-zero. After specifically utilizing a GHZ channel and measurement quantum teleportation protocol, a general proposition extending the work carried out in Ref. \citen{Barasinski_2018} was derived showing that localizable concurrence is necessary, not only for controlled quantum teleportation protocols, but for any quantum teleportation protocol of a single qubit relying on a Bell measurement. Our work stresses the importance of localizable concurrence as a quantum resource for teleportation protocols via multi-partite entangled states, calling for more investigation on its quantification and relation to multi-partite entanglement. Practical scenarios where the localizable concurrence has to be quantified include $n$-party collaborative teleportation protocols, where up to $n-2$ agents have to perform optimal local measurements on their qubits in order to allow, possibly not predetermined, $n$-th agent to successfully complete the protocol.

\section*{Acknowledgments}
We acknowledge fruitful discussions with Quantumalta group members André Xuereb and Jake Xuereb.

\bibliographystyle{ws-ijqi}
\bibliography{biblio}

\begin{thebibliography}{10}

\bibitem{Plenio_2007}
M.~B. Plenio and S.~Virmani, {\em Quantum Inf. Comput.} {\bf 7}  (2007) 1.

\bibitem{Steane_1998}
A.~Steane, {\em Reports on Progress in Physics} {\bf 61} (Feb 1998) p.
  117–173.

\bibitem{Gisin_2007}
N.~Gisin and R.~Thew, {\em Nature Photonics} {\bf 1} (Mar 2007) p. 165–171.

\bibitem{Bennett_2014}
C.~H. Bennett and G.~Brassard, {\em Theoretical Computer Science} {\bf 560}
  (Dec 2014) p. 7–11.

\bibitem{Bennett_1993}
C.~H. Bennett, G.~Brassard, C.~Cr{\'{e}}peau, R.~Jozsa, A.~Peres and W.~K.
  Wootters, {\em Phys. Rev. Lett.} {\bf 70}  (1993) 1895.

\bibitem{Pirandola2015}
S.~Pirandola, J.~Eisert, C.~Weedbrook, A.~Furusawa and S.~Braunstein, {\em
  Nature Photonics} {\bf 9} (05 2015)

\bibitem{Jozsa_1994}
R.~Jozsa, {\em J. Mod. Opt.} {\bf 41}  (1994) 2315.

\bibitem{Popescu_1994}
S.~Popescu, {\em Physical review letters} {\bf 72} (February 1994) p.
  797—799.

\bibitem{Massar_1995}
S.~Massar and S.~Popescu, {\em Phys. Rev. Lett.} {\bf 74} (Feb 1995) 1259.

\bibitem{Verstraete_2002}
F.~Verstraete and H.~Verschelde, {\em Physical Review A} {\bf 66} (Aug 2002)

\bibitem{Wootters_1998}
W.~K. Wootters, {\em Phys. Rev. Lett.} {\bf 80}  (1998) 2245.

\bibitem{Horodecki2009}
R.~Horodecki, P.~Horodecki, M.~Horodecki and K.~Horodecki, {\em Reviews of
  Modern Physics} {\bf 81} (jun 2009) 865.

\bibitem{Guhne2009}
O.~G{\"{u}}hne and G.~T{\'{o}}th, {\em Physics Reports} {\bf 474}  (2009) 1.

\bibitem{Karlsson_1998}
A.~Karlsson and M.~Bourennane, {\em Phys. Rev. A} {\bf 58} (Dec 1998) 4394.

\bibitem{Lee2005}
S.~Lee, J.~Joo and J.~Kim, {\em Physical Review A - Atomic, Molecular, and
  Optical Physics} {\bf 72}  (2005) 2.

\bibitem{Agrawal_2006}
P.~Agrawal and A.~Pati, {\em Physical Review A} {\bf 74} (Dec 2006)

\bibitem{Yang2009}
K.~Yang, L.~Huang, W.~Yang and F.~Song, {\em International Journal of
  Theoretical Physics} {\bf 48} (feb 2009) 516.

\bibitem{Dur2000}
W.~Dur, G.~Vidal and J.~I. Cirac, {\em Phys. Rev. A} {\bf 62}  (2000)

\bibitem{Ac_n_2001}
A.~Acín, D.~Bruß, M.~Lewenstein and A.~Sanpera, {\em Physical Review Letters}
  {\bf 87} (Jul 2001)

\bibitem{greenberger2007}
D.~M. Greenberger, M.~A. Horne and A.~Zeilinger, Going beyond bell's theorem
  (2007).

\bibitem{Bose1998}
S.~Bose, V.~Vedral and P.~L. Knight, {\em Phys. Rev. A} {\bf 57} (Feb 1998) p.
  822–829.

\bibitem{QuantumInformation}
S.~Barnett, {\em Quantum Information} (Oxford University Press, Inc., USA,
  2009).

\bibitem{Moreno_2018}
M.~G.~M. Moreno, A.~Fonseca and M.~M. Cunha, {\em Quantum Information
  Processing} {\bf 17} (Jun 2018)

\bibitem{Joo2003}
J.~Joo, Y.~J. Park, S.~Oh and J.~Kim, {\em New Journal of Physics} {\bf 5}
  (2003)

\bibitem{Barasinski_2018}
A.~Barasiński, I.~Arkhipov and J.~Svozilík, {\em Scientific Reports} {\bf 8}
  (12 2018)

\bibitem{Verstraete_2004}
F.~Verstraete, M.~Popp and J.~I. Cirac, {\em Physical Review Letters} {\bf 92}
  (Jan 2004)

\bibitem{Popp_2005}
M.~Popp, F.~Verstraete, M.~A. Martín-Delgado and J.~I. Cirac, {\em Physical
  Review A} {\bf 71} (Apr 2005)

\bibitem{Eltschka_2014}
C.~Eltschka and J.~Siewert, {\em Journal of Physics A: Mathematical and
  Theoretical} {\bf 47} (Oct 2014) p. 424005.

\bibitem{Cunha_2019}
M.~M. Cunha, A.~Fonseca and E.~O. Silva, {\em Universe} {\bf 5} (Oct 2019) p.
  209.

\bibitem{Eltschka_2012}
C.~Eltschka and J.~Siewert, {\em Physical Review Letters} {\bf 108} (Jan 2012)

\bibitem{Coffman_2000}
V.~Coffman, J.~Kundu and W.~K. Wootters, {\em Physical Review A} {\bf 61} (Apr
  2000)

\bibitem{D_r_1999}
W.~Dür, J.~I. Cirac and R.~Tarrach, {\em Physical Review Letters} {\bf 83}
  (Oct 1999) p. 3562–3565.

\bibitem{de_Vicente_2011}
J.~I. de~Vicente and M.~Huber, {\em Physical Review A} {\bf 84} (Dec 2011)

\bibitem{Huber_2014}
M.~Huber and R.~Sengupta, {\em Phys. Rev. Lett.} {\bf 113} (Sep 2014) p.
  100501.

\bibitem{Horodecki_1998}
K.~Zyczkowski, P.~Horodecki, A.~Sanpera and M.~Lewenstein, {\em Physical Review
  A - PHYS REV A} {\bf 58} (08 1998) 883.

\bibitem{Jung2008}
E.~Jung, M.~R. Hwang, Y.~H. Ju, M.~S. Kim, S.~K. Yoo, H.~Kim, D.~Park, J.~W.
  Son, S.~Tamaryan and S.~K. Cha, {\em Physical Review A - Atomic, Molecular,
  and Optical Physics} {\bf 78}  (2008)

\bibitem{Li2010a}
Y.~L. Li, M.~F. Fang, X.~Xiao, C.~Wu and L.~Z. Hou, {\em Chinese Physics B}
  {\bf 19}  (2010) 1.

\bibitem{Chun_2010}
M.~Chun, Y.~Ming and C.~Zhuo-Liang, {\em Communications in Theoretical Physics}
  {\bf 53} (mar 2010) 489.

\bibitem{Hu2011}
M.~L. Hu, {\em Physics Letters, Section A: General, Atomic and Solid State
  Physics} {\bf 375}  (2011) 922.

\bibitem{Liang2015}
H.~Q. Liang, J.~M. Liu, S.~S. Feng, J.~G. Chen and X.~Y. Xu, {\em Quantum
  Information Processing} {\bf 14}  (2015) 3857.

\bibitem{NC2011}
M.~A. Nielsen and I.~L. Chuang, {\em Quantum Computation and Quantum
  Information: 10th Anniversary Edition}, 10th edn. (Cambridge University
  Press, USA, 2011).

\bibitem{Li_2007}
L.~Li and D.~Qiu, {\em Journal of Physics A: Mathematical and Theoretical} {\bf
  40} (Aug 2007) p. 10871–10885.

\bibitem{Hillery1999}
M.~Hillery, V.~Bu{\v{z}}ek and A.~Berthiaume, {\em Phys. Rev. A - At. Mol. Opt.
  Phys.} {\bf 59}  (1999) 1829.

\bibitem{Man2007}
Z.~X. Man, Y.~J. Xia and N.~B. An, {\em Eur. Phys. J. D} {\bf 42}  (2007) 333.

\bibitem{Hao2001}
J.~C. Hao, C.~F. Li and G.~C. Guo, {\em Phys. Rev. A - At. Mol. Opt. Phys.}
  {\bf 63}  (2001) p.~3.

\bibitem{Komar2014}
P.~K{\'{o}}m{\'{a}}r, E.~M. Kessler, M.~Bishof, L.~Jiang, A.~S. S{\o}rensen,
  J.~Ye and M.~D. Lukin, {\em Nat. Phys.} {\bf 10}  (2014) 582.

\bibitem{Roos2004}
C.~F. Roos, M.~Riebe, H.~H{\"{a}}ffner, W.~H{\"{a}}nsel, J.~Benhelm, G.~P.
  Lancaster, C.~Becher, F.~Schmidt-Kaler and R.~Blatt, {\em Science (80-. ).}
  {\bf 304}  (2004) 1478.

\bibitem{DeLimaBernardo2019}
B.~{de Lima Bernardo}, M.~Lencses, S.~Brito and A.~Canabarro, {\em Quantum Inf.
  Process.} {\bf 18}  (2019) 1.

\bibitem{Miao2010}
C.~Miao, M.~Yang and Z.~L. Cao, {\em Commun. Theor. Phys.} {\bf 53}  (2010)
  489.

\bibitem{Yang2019}
Y.~G. Yang, S.~Gao, D.~Li, Y.~H. Zhou and W.~M. Shi, {\em Quantum Inf.
  Process.} {\bf 18}  (2019) 1.

\bibitem{Gorbachev2000}
V.~N. Gorbachev and A.~I. Trubilko, {\em Journal of Experimental and
  Theoretical Physics} {\bf 91} (nov 2000) 894.

\end{thebibliography}

\end{document}